\documentclass[aps,prd,secnumarabic,nobibnotes,twocolumn,superscriptaddress]{revtex4-1}
\usepackage{amsfonts}
\usepackage{mathrsfs}
\usepackage{amsmath}
\usepackage{color}
\usepackage{natbib}
\usepackage{graphicx}
\usepackage{bm}
\usepackage{amssymb}
\usepackage{xspace}
\usepackage{epstopdf}
\usepackage{dcolumn}
\usepackage{multirow}
\usepackage[colorlinks=true, letterpaper=true, pdfstartview=FitV, linkcolor=blue, citecolor=blue, urlcolor=blue]{hyperref}
\usepackage{wrapfig}

\makeatletter

\newcommand{\Rmnum}[1]{\expandafter\@slowromancap\romannumeral #1@}
\makeatother

\begin{document}
\title{Two-dimensional Weyl nodal line semimetal in high Curie temperature $d^{0}$ ferromagnet K$_2$N monolayer}

\author{Lei Jin}
\affiliation{School of Materials Science and Engineering, Hebei University of Technology, Tianjin 300130, China.}

\author{Xiaoming Zhang}
\email{zhangxiaoming87@hebut.edu.cn}
\affiliation{School of Materials Science and Engineering, Hebei University of Technology, Tianjin 300130, China.}

\author{Ying Liu}
\address{Research Laboratory for Quantum Materials, Singapore University of Technology and Design, Singapore 487372, Singapore}

\author{Xuefang Dai}
\affiliation{School of Materials Science and Engineering, Hebei University of Technology, Tianjin 300130, China.}

\author{Xunan Shen}
\affiliation{School of Materials Science and Engineering, Hebei University of Technology, Tianjin 300130, China.}

\author{Liying Wang}
\affiliation{Tianjin Key Laboratory of Low Dimensional Materials Physics and Preparation Technology, School of science, Tianjin University, Tianjin 300354, People's Republic of China.}

\author{Guodong Liu}
\email{gdliu1978@126.com}
\affiliation{School of Materials Science and Engineering, Hebei University of Technology, Tianjin 300130, China.}

\begin{abstract}
Nodal line semimetals in two-dimensional (2-D) materials have attracted intense attention currently. From fundamental physics and spintronic applications points of view, high Curie temperature ferromagnetic (FM) ones with nodal lines robust against spin-orbit coupling (SOC) are significantly in desirable. Here, we propose that FM K$_2$N monolayer is such Weyl nodal line semimetal. We show that K$_2$N monolayer is dynamically stable, and has a FM ground magnetic state with the out-of-plane [001] magnetization. It shows two nodal lines in the low-energy band structures. Both nodal lines are robust against SOC, under the protection of mirror symmetry. We construct an effective Hamiltonian, which can well characterize the nodal lines in the system. Remarkably, the nodal line semimetal proposed here is distinct from the previously studied ones in that K$_2$N monolayer is 2-D $d^{0}$-type ferromagnet with the magnetism arising from the partially filled N-$p$ orbitals, which can bring special advantages in spintronic applications. Besides, the Curie temperature in K$_2$N monolayer is estimated to be 942K, being significantly higher than previous FM nodal lines materials. We also find that, specific tensile strains can transform the nodal line from type-I to a type-II one, making its nodal line characteristics even more interesting.
\end{abstract}
\maketitle

\section{Introduction}
Topological nodal line semimetals/metals (TNLSMs) have attracted tremendous research interests recently. They are characterized with nontrivial nodal line state in the Brillouin zone (BZ), formed by the one-dimensional band crossings in the low-energy band structures~\cite{1,2,3,4}. Nodal lines can have different forms of band crossings. For one example, the band crossings for nodal lines can show both the fourfold degeneracy and the double degeneracy, and the nodal lines are known as Dirac and Weyl nodal lines~\cite{5,6,7,8,9}. For the latter, it requires the inversion symmetry or the time reversal symmetry is broken. For another example, band crossings for nodal lines can show different slopes of band dispersions namely type-I or type-II band crossings. When band crossings for a nodal line are all type-I (or type-II), the nodal line corresponds to a type-I (or type-II) nodal line~\cite{10,11,12}. If a nodal line possesses both type-I and type-II band crossings, it belongs to a hybrid type~\cite{13,14}. TNLSMs show many interesting electron properties such as the drumhead surface state~\cite{1,15}, novel electrical/optical response~\cite{16,17}, and rich transport characteristics~\cite{18,19,20,21}, and have been well studied in three-dimensional (3-D) materials both theoretically and experimentally~\cite{22,23,24,25,26,27,28}.

In two-dimensional (2-D) materials, several TNLSMs have also been predicted in theory~\cite{29,30,31,32,33,34,35}. Remarkably, the nodal line state in 2-D Cu$_2$Si~\cite{36} and CuSe~\cite{37} have already been verified by experiments recently. These 2-D TNLSMs are all nonmagnetic (NM) examples. Recently, special efforts have been made to explore ferromagnetic (FM) 2-D nodal lines, because the spin vector is liberated in FM materials and can favor potential applications in spintronics. To date, only a few FM 2-D TNLSMs were reported, including layered Fe$_2$Sn~\cite{38}, GdAg$_2$ monolayer~\cite{39}, MnN monolayer~\cite{40}, layered Fe$_3$GeTe$_2$~\cite{41}, Na$_2$CrBi trilayer~\cite{42}, and XCl (X= Gd, Sc, Y, La) monolayer~\cite{43,44,45,46}. These examples have confirmed the feasibility of realizing TNLSMs in FM 2-D materials, but they still face two major issues. The first issue is that, most nodal lines in these FM 2-D materials will be gapped under spin-orbit coupling (SOC). Especially, the SOC gaps in Fe$_3$Ge$_2$Te$_2$ and GdBr can be as large as 60 meV~\cite{41,43}. MnN monolayer is the only example with FM nodal lines robust against SOC~\cite{40}. The second issue is the relatively low Curie temperature in existing FM 2-D TNLSMs. For example, the Curie temperature in MnN monolayer is about 200 K~\cite{40}, lower than room the temperature. Therefore, exploring high Curie temperature FM 2-D TNLSMs with nodal lines robust against SOC is urgently needed.

In addition, in existing FM 2-D TNLSMs, the magnetic moments are mainly contributed by the unoccupied $d$/$f$ shells from the transition metal or rare earth elements which are known as $d$/$f$-type ferromagnetism. Recently, FM 2-D materials in the absence of unoccupied $d$/$f$ shells, namely \emph{d$^{0}$} ferromagnets, were reported in a few 2-D materials, such as Na$_2$C monolayer~\cite{47}, C$_3$Ca$_2$ sheet~\cite{48}, and YN$_2$ monolayer~\cite{49}. The magnetisms in these examples arise from the $p$ orbitals. Compared to the $d$/$f$-type counterpart, \emph{d$^{0}$} ferromagnets are more likely to show higher carrier mobility and longer spin coherence length because of the much weaker localization of \emph{p} electrons and smaller SOC strength, which are significant advantages for high-speed and long-distance transports~\cite{47,48,49}. Then one may wonder that, can 2-D nodal line semimetal exist in \emph{$d^{0}$} ferromagnets?

In current work, based on electronic structure calculations and symmetry analysis, we propose K$_2$N monolayer as the first candidate of 2-D nodal line material in \emph{$d^{0}$} ferromagnets, with the presence of nodal lines robust against SOC. K$_2$N monolayer, as a member of 2-D AXenes, was initially proposed by Zhao \emph{et al}. under systematic material screening~\cite{50}. They have demonstrated the material is dynamically and thermally stable, and has a robust ferromagnetism with long-range $p$ state coupling. Here, we reveal that K$_2$N monolayer shows two nodal lines in its low-energy band structures. The nodal lines are concentric at the \emph{$\Gamma$} point in the 2-D Brillouin zone. Remarkably, both nodal lines are robust against SOC, because they are formed by bands with opposite $M$$_z$ eigenvalues ($\pm$$i$). We have constructed an effective Hamiltonian to describe the topological nature of the nodal lines. Moreover, we find the nodal lines can be successively annihilated under tensile in-plane strains. At specific strains, one of the nodal lines can transform from type-I to a type-II one, making the nodal line states of K$_2$N monolayer even interesting. Considering the extremely high Curie temperature (942 K) and the clean nodal line band structure, K$_2$N monolayer can plays a good material platform to investigate the fundamental physics of 2-D nodal-line fermions in \emph{$d^{0}$} ferromagnets.

\section{COMPUTATIONAL METHODS}
The first-principles calculations in this work are carried out by using the Vienna ab initio Simulation Package (VASP)~\cite{51}, based on density functional theory (DFT)~\cite{52}. For the crystal structure of K$_2$N monolayer, we build a large vacuum space of 15 {\AA} to avoid potential interactions between layers. During the calculations, the long-range van der Waals interactions are taken into account by using the DFT-D2 method~\cite{53}. The cutoff energy is set as 700 eV. The Brillouin zone is sampled by a Monkhorst-Pack $k$-mesh with size of 13$\times$13$\times$1. In the calculations, the force and energy convergence criteria are set as 0.01 eV${\AA}^{-1}$ and $10^{-6}$ eV, respectively. The irreducible representations of the electronic states are obtained by using the irvsp code~\cite{54}. To investigate the dynamical stability of K$_2$N monolayer, the phonon spectra are calculated by using the PHONOPY code~\cite{55}. The edge states are calculated by using the WANNIERTOOLS package~\cite{56}.

\section{Structure and stability}
\begin{figure}
\includegraphics[width=8.8cm]{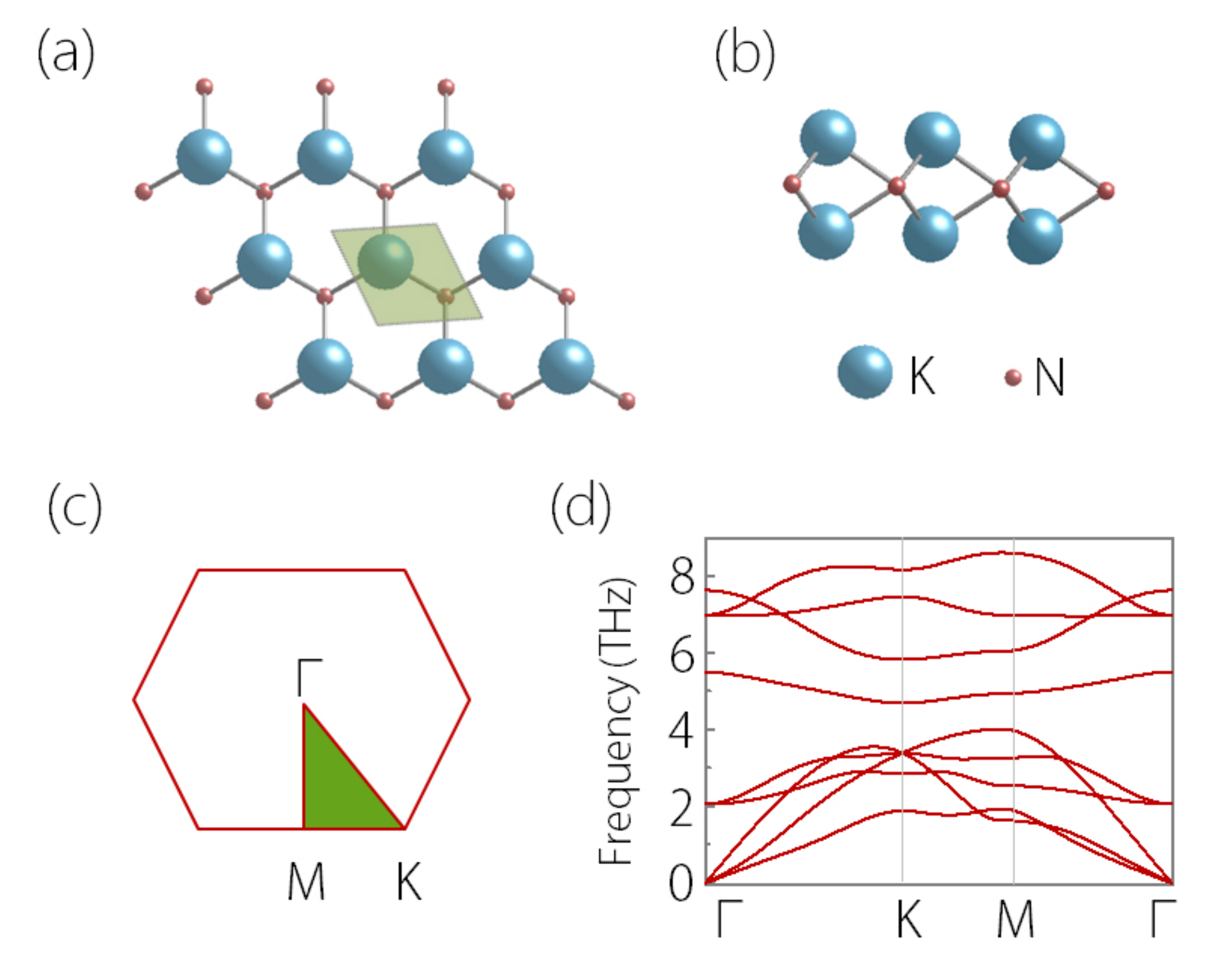}
\caption{(a) Top and (b) side view of crystal structure for K$_2$N monolayer. The shadowed region in (a) shows its primitive cell. (c) The 2-D Brillouin zone of K$_2$N monolayer. (d) The calculated phonon spectrum of K$_2$N monolayer.
\label{fig1}}
\end{figure}

The lattice structure of K$_2$N monolayer is shown in Fig.~\ref{fig1}(a) and (b). It exhibits a hexagonal structure with the space group \emph{P\={6}M2} (No. 187), and the point group is $D$$_{3h}$. In the side view [see Fig.~\ref{fig1}(b)], the unit cell of K$_2$N contains two K atomic layers and one N layer, and their bonding forms the triple-layered K-N-K sandwich structure. Specifically, in the top view [see Fig.~\ref{fig1}(a)], we can observe a honeycomb structure, as formed by the bonding between N and K atoms. The shadowed region in Fig.~\ref{fig1}(a) shows the primitive cell of K$_2$N monolayer. One primitive cell contains one unit of K$_2$N formula. The lattice constant for K$_2$N monolayer is optimized as a = b = 4.058 {\AA}, and the K-N bond length is 2.822 {\AA}. These values are consistent with the previously reported ones~\cite{50}.

Here we check the stability of the obtained K$_2$N monolayer. We first estimate the thermal stability of K$_2$N monolayer by calculating the formation energy ($H$$_f$) following the equation:

\begin{equation}\label{FNRm}
H_{f}=(2E_{K}+E_{N}-E_{K_{2}N})/3
\end{equation}
where \emph{E$_K$}, \emph{E$_N$}, and \emph{E$_{K_{2}N}$} are the total energies of a single K atom in pure metal, a single N atom in nitrogen, and one unit cell of K$_2$N, respectively. The obtained formation energy of K$_2$N monolayer is 2.19 eV/atom, which is larger than that of 2-D Ca$_2$C$_3$ (1.38 eV/atom)~\cite{48}, VN monolayer (1.47-1.68 eV/atom)~\cite{57} and Na$_2$C monolayer (1.38 eV/atom)~\cite{47}. This suggests K$_2$N monolayer is thermally stable.

To confirm the dynamic stability of K$_2$N monolayer, we have calculated the phonon spectra and the results are shown in Fig.~\ref{fig1}(d). In the phonon spectra, the optical and acoustical branches are well separated with each other, and exhibit no imaginary mode throughout all high symmetry $k$-paths in the 2-D Brillouin zone. Moreover, near the $\Gamma$ point, we can observe that in-plane acoustic phonons show the linear dispersion, while out-plane acoustic branch exhibits the quadratic dispersion. This is a typical signature for 2-D materials~\cite{58,59}. In this respect, the given structure of K$_2$N monolayer has an excellent dynamical stability.

\section{Magnetic property}
\begin{figure}
\includegraphics[width=8.8cm]{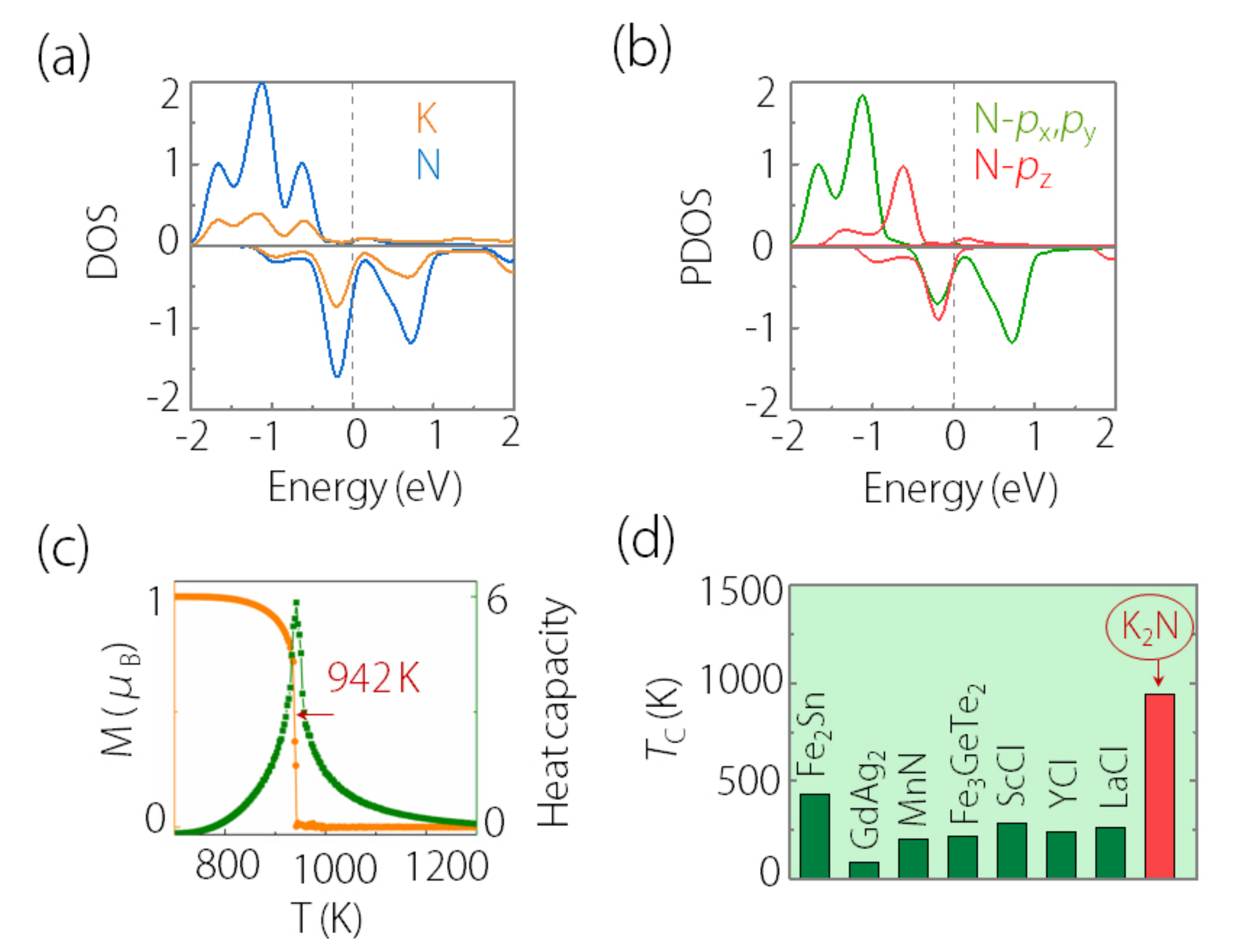}
\caption{(a) The total density of state (DOS)of K$_2$N monolayer. (b) The projected density of state (PDOS)of K$_2$N monolayer. (c) The magnetic moment ($M$) and specific heat capacity ($C$$_V$) magnetic susceptibility as a function of temperature. (d) The Curie temperature of K$_2$N monolayer compared with the values of the previously reported FM 2-D nodal line materials. Some data are obtained from literatures~\cite{38,39,40,41,46}.
\label{fig2}}
\end{figure}

We notice K$_2$N monolayer has an unconventional stoichiometry, which may introduce magnetism in the system. This has been verified by our DFT calculations. In our calculations, we have considered all potential magnetic configurations including NM, FM, and antiferromagnetic (AFM). We compare their energies to determine the ground magnetic state of K$_2$N monolayer. Our calculations show the FM state has the lowest energy, which is 0.09 eV lower than the AFM state and 0.31 eV lower than the NM state for one unit cell. Therefore, the ground magnetic state of K$_2$N monolayer is FM. In the FM state, we find the magnetic moment in K$_2$N monolayer mainly originates from the N atoms, more specifically, from the $p$$_x$/$p$$_y$ orbitals near the Fermi level. As shown by the total and projected density of states (DOSs) in Fig.~\ref{fig2}(a) and (b),we can observe that the N-$p$$_x$/$p$$_y$ orbitals are degenerate and partially filled in both spin channels, while the N-$p$$_z$ orbitals are fully occupied. The magnetic moment on the N atom is 0.872 $\mu$$_B$, and the total magnetic moment of K$_2$N monolayer is 0.983 $\mu$$_B$. These results confirm that K$_2$N monolayer is a typical \emph{$d^{0}$} ferromagnet.

To determine the easy axis for the FM state, we have calculated the magnetic anisotropy energy along different magnetization directions with SOC taken into account. Our calculations show that the energies at different magnetization directions are almost degenerate, while the out-of-plane [001] magnetization has a bit lower energy ($< 0.1$ meV) than others. These results indicate K$_2$N monolayer has a very "soft" magnetism, and the ground magnetic state is FM with magnetization along the out-of-plane [001] direction.

The Curie temperature, which determines the thermal stability of the FM ordering, is very crucial for practical applications in spintronic devices. Here we perform the Monte Carlo simulation to estimate the Curie temperature of K$_2$N monolayer. The simulation is based on the Ising model~\cite{60}. The Hamiltonian of classical Heisenberg model can be expressed as:

\begin{equation}\label{FNRm}
H=-\Sigma_{ij}J_{ij}M_{i}M_{j}
\end{equation}
where \emph{J$_{ij}$} is the nearest-neighboring exchange parameter, and $M$ represents the magnetic moment on the N site. To weaken the periodic constraints, a 100$\times$100 supercell of K$_2$N monolayer is used here. Beside the magnetic moment, we have also calculated the temperature-dependent heat capacity ($C$$_V$) by using the equation:
\begin{equation}\label{FNRm}
C_{V}=\frac{\langle E^{2}\rangle-\langle E\rangle^{2}}{\langle L^{2}T^{2}\rangle}
\end{equation}
where $E$ is the energy corresponding to each magnetic moment. In Fig.~\ref{fig2}(c), we show the curves of average magnetic moment and heat capacity versus the temperature. Excitingly, the Curie temperature of K$_2$N monolayer can be as high as 942 K. The high temperature ferromagnetism makes K$_2$N monolayer is promising for practical spintronic applications.

\section{Robust weyl nodal line}
\begin{figure}
\includegraphics[width=8.8cm]{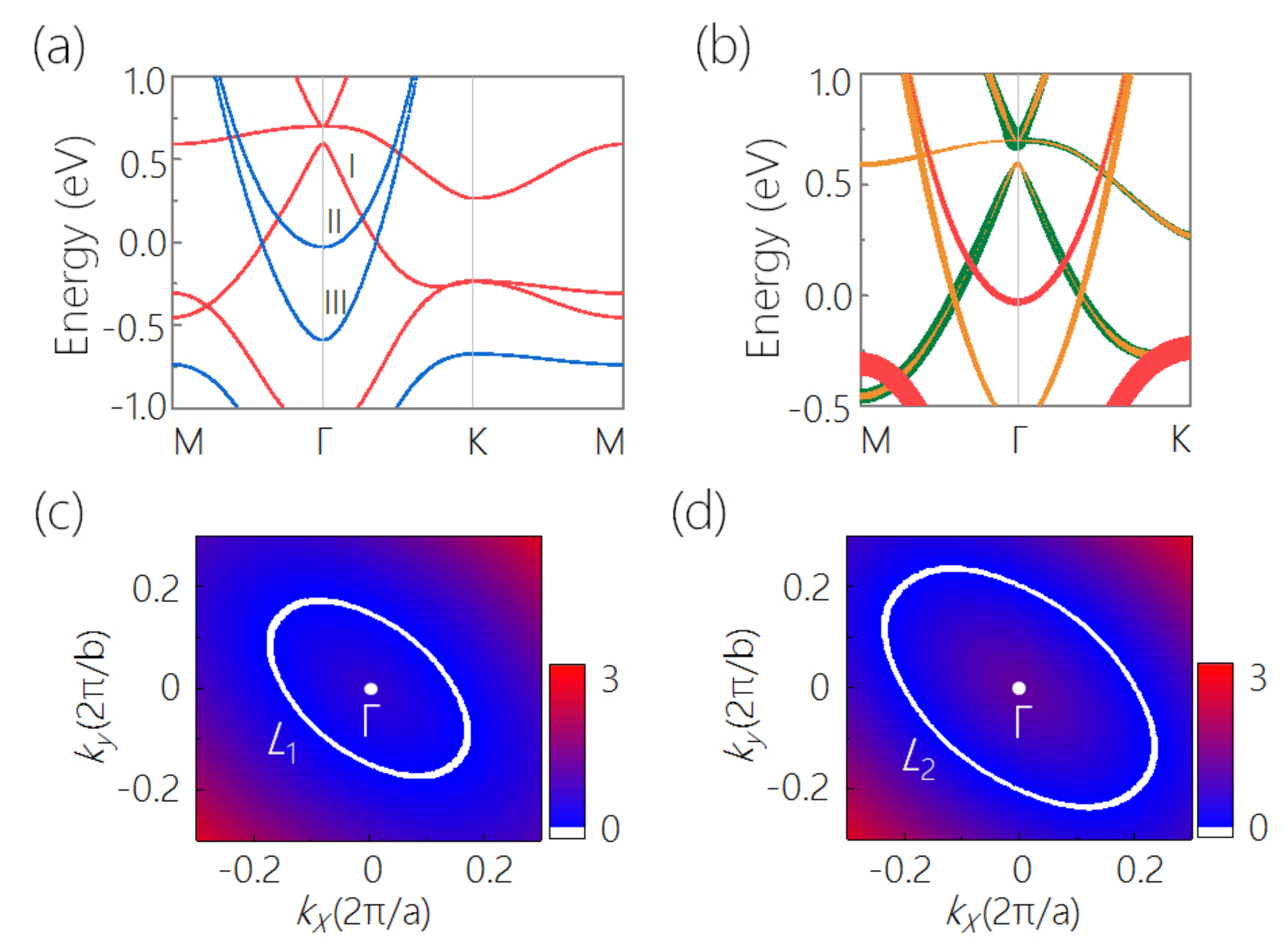}
\caption{(a) The spin-polarized band structure of K$_2$N monolayer. The blue and red lines denote the spin-up and spin-down channels, respectively. The three crossing bands near the Fermi level are labeled as bands I, II and III, respectively. (b) The orbital-projected band structure of K$_2$N monolayer, including N-$p$$_y$ orbital (green), N-$p$$_z$ orbital (red), and K-$p$ orbitals (yellow). (c) and (d) show the shapes of two nodal lines in K$_2$N monolayer, which are labeled as $L$$_1$ and $L$$_2$, respectively.
\label{fig3}}
\end{figure}

Here we investigate the topological band structure of K$_2$N monolayer. We first discuss the case without considering SOC. The spin-resolved band structure is shown in Fig.~\ref{fig3}(a). We can find that K$_2$N monolayer exhibits a metallic band structure with three bands crossing the Fermi level. Among the three low-energy bands, one is hole-like and arises from the spin-down channel (labeled as band I); and the other two are electron-like and arise from the spin-up channel (labeled as bands II and III). Band I crosses with bands II and III, forming four band crossing points in the $M$-$\Gamma$-$K$ path. Further orbital-component analysis shows that these bands are mostly contributed by the $p$$_y$ (green) and $p$$_z$ (red) orbitals of N atom and $p$ (yellow) orbitals of K atom, as shown in Fig.~\ref{fig3}(b). From the orbital-projected bands, we can clearly observe the signature of band inversions at the $\Gamma$ point, indicating the potential nontrivial band topology in K$_2$N monolayer.

By a careful scan of the band structures in the 2-D Brillouin zone, we find the crossings among the three bands in fact form two concentric nodal lines at the $\Gamma$ point, as shown in Fig.~\ref{fig3}(c) and (d). The nodal line formed by bands I and II is smaller in size, and we label it as $L$$_1$. The other nodal line (labeled as $L$$_2$) is formed by bands I and III, and has a bigger size. The profiles of $L$$_1$ and $L$$_2$ are shown in Fig.~\ref{fig3}(c) and (d), respectively. To be noted, both nodal lines $L$$_1$ and $L$$_2$ locate close to the Fermi level, which greatly favors their detections in experiments.

\begin{figure}
\includegraphics[width=8.8cm]{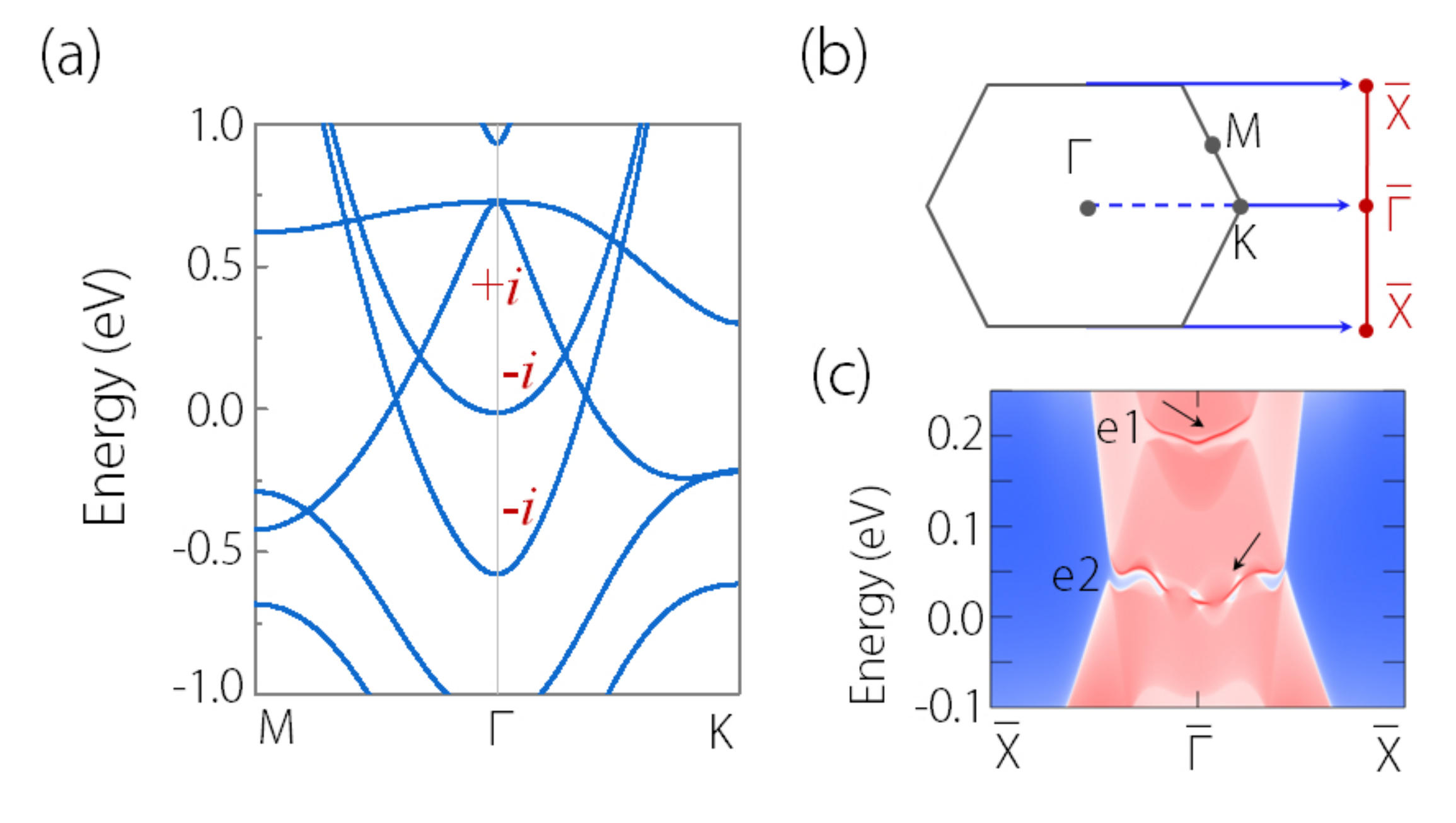}
\caption{(a) The electronic band structure of K$_2$N monolayer with SOC under out-of-plane [001] magnetization. The $M$$_z$ eigenvalues (+$i$ or -$i$) are labeled for bands I, II, and III. (b) The 2-D Brillouin zone of K$_2$N monolayer and its projection onto the (010) edge. (c) The Projected spectrum on the (010) edge. The e1 and e2 correspond to the edge states for nodal lines $L$$_1$ and $L$$_2$, respectively.
\label{fig4}}
\end{figure}

Then we turn on SOC in the calculations. The enlarged band structure along the $M$-$\Gamma$-$K$ path under SOC is shown in Fig.~\ref{fig4}(a). Very interestingly, we find the band crossings in the $M$-$\Gamma$-$K$ path are not gapped under SOC. We have also checked other parts on both nodal lines, and still find no gaps. These results show the nodal lines in K$_2$N monolayer are robust against SOC. Then one may naturally wonder, what protects the nodal lines in K$_2$N monolayer under SOC? By checking the symmetry signatures of these low-energy bands, we find that the two nodal lines are in fact protected by the horizontal mirror symmetry $M$$_z$. As we have confirmed above, the magnetization in K$_2$N monolayer is along the out-of-plane [001] direction. Such a magnetization direction will not break $M$$_z$ in the FM state. Our symmetry analysis finds the hole-like band I possesses the $M$$_z$ eigenvalue of +\emph{i}, while the two electron-like bands II and III possess the opposite $M$$_z$ eigenvalue of -\emph{i}, as indicated in Fig.~\ref{fig4}(a). The opposite $M$$_z$ eigenvalues between band I and band II (or band III) require they cross with each other without hybridization. As a result, the two nodal lines will retain under SOC as long as the mirror symmetry is preserved.

In 3-D nodal line semimetals, nodal lines are characterized with nontrivial drumhead surface states~\cite{1,26,61}. In 2-D, nodal lines may also show nontrivial edge states. We have calculated the band spectrum of the (010) edge for K$_2$N monolayer. The results are shown in Fig.~\ref{fig4}(c). Based on the energy range of the nodal lines, the edge states e1 and e2 correspond to the nodal lines $L$$_1$ and $L$$_2$, respectively. The appearance of these edge states may be an indicator of the nontrivial band topology in K$_2$N monolayer.

To character the nature of both nodal lines, we have constructed an effective \emph{k$\cdot$p} model for the three crossing bands. When SOC is included, at the $\Gamma$ point, band I belongs to the $\Gamma$$_3$ irreducible representation of the $D$$_{4h}$ point group symmetry, whereas bands II and III both belong to the $\Gamma$$_4$ irreducible representation. Three symmetries including the mirror symmetry $M$$_z$, the threefold rotation symmetry $C$$_{3z}$, and the twofold rotation symmetry $C$$_{2x}$ should be considered in the model. Using the states of bands I and II as the basis, the effective Hamiltonian for $L$$_1$ (up to the quadratic order) can be expressed as:
\begin{equation}\label{FNRm}
H(k_{x},k_{y})=\left[
              \begin{array}{cc}
               (m+\alpha k^{2}) & 0\\
0 & -(m+\alpha k^{2})
              \end{array}
            \right]
\end{equation}
where $k^2$ =\emph{k$_x^2$}+\emph{k$_y^2$}  and the coefficients \emph{m} and $\alpha$ are real and material-dependent parameters. Their values can be obtained by fitting the DFT band structure. In (4), the nodal line only appears when \emph{m*$\alpha$}$<0$. After fitting the DFT band structure, we indeed find \emph{m*$\alpha$}$<0$ for $L$$_1$. To be noted, the effective Hamiltonian for $L$$_2$ has the same form of (4), because bands II and III have the same irreducible representation.

\begin{figure}
\includegraphics[width=8.8cm]{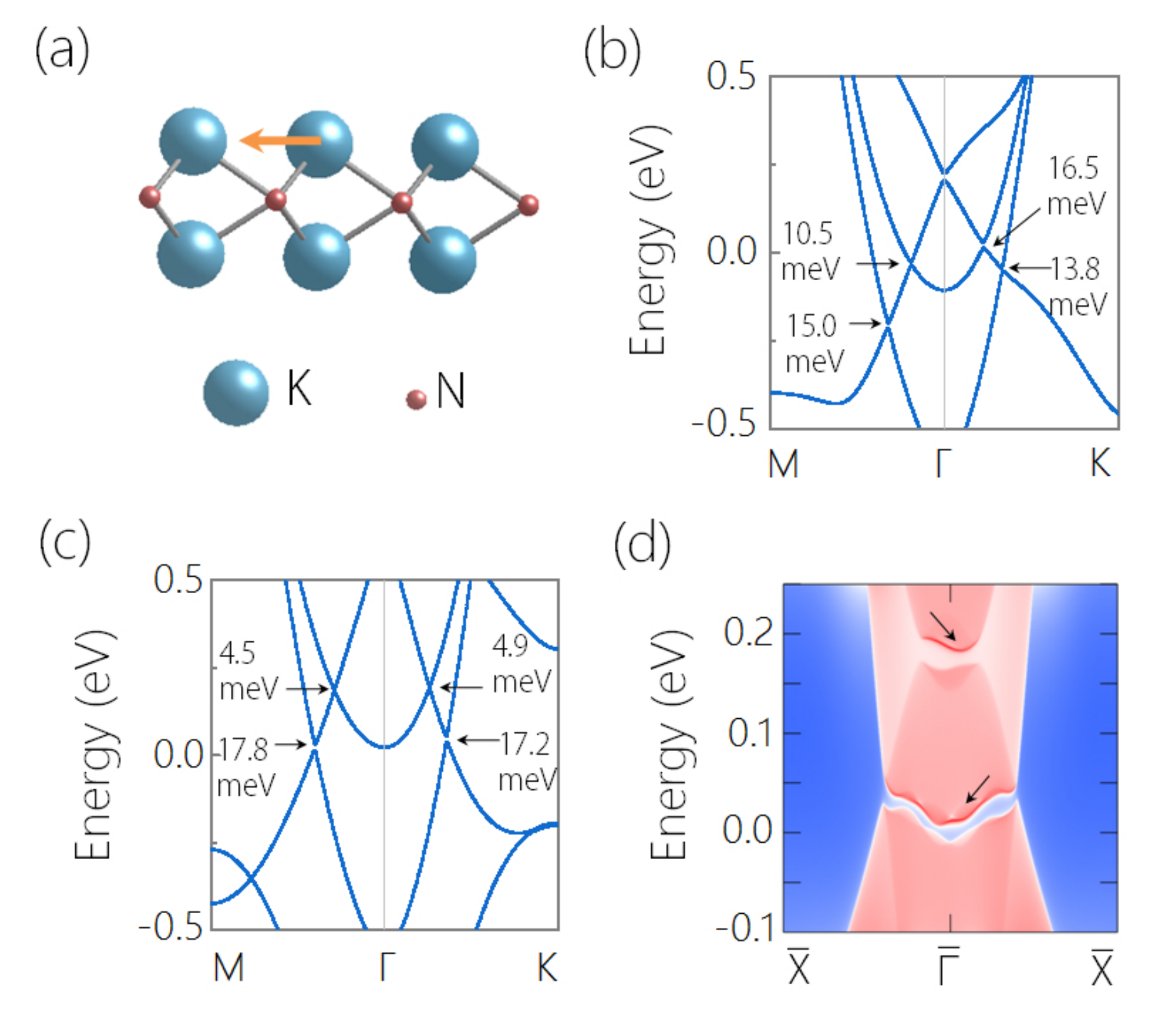}
\caption{(a) The lattice distortion of K$_2$N monolayer by artificially shifting one of the upper K atoms towards the neighbouring N atom, which will break the mirror symmetry $M$$_z$. (b) The corresponding band structure with SOC for (a). (c) The band structure (considering SOC) of K$_2$N monolayer with in-plane [100] magnetization. (d) Corresponding (010) edge states for (c). In (b) and (c), the sizes of SOC gaps are indicated.
\label{fig5}}
\end{figure}

In above discussions, the mirror symmetry $M$$_z$ is preserved after considering SOC, which protects the nodal lines in K$_2$N monolayer. Breaking $M$$_z$ will generally remove the nodal lines. To show this point, we here break the $M$$_z$ by two ways. First, we horizontally shift one of the upper K atoms towards the neighbouring N atom, as shown in Figs.~\ref{fig5}(a). Corresponding band structure is shown in Figs.~\ref{fig5}(b). We observe the nodal lines are indeed gapped, where the gap size along the $M$-$\Gamma$-$K$ path is in the range of 10-17 meV. Second, we shift the magnetization direction into in-plane, which would break the $M$$_z$. As shown in Figs.~\ref{fig5}(c), the nodal lines are again gapped out with the gap size of 4.5-18 meV. Figs.~\ref{fig5}(d) shows the edge states in this case.

\section{Strain-induced nodal line transformation}
Based on the slopes of the crossing bands, nodal lines can be classified into type-I, type-II, and the hybrid type~\cite{13,14}. For type-I (type-II) nodal lines, the band crossings are purely type-I (II)~\cite{13}, whereas hybrid nodal lines can simultaneously show both type-I and type-II band crossings, realized by a saddle-shape band~\cite{14}.  In the native state, nodal lines $L$$_1$ and $L$$_2$ in K$_2$N monolayer both belong to type-I nodal lines, since they each is formed by one hole-like band and one electron-like band. Very interestingly, we find both the number and the type of nodal lines can be shifted by simply assigning lattice strains on K$_2$N monolayer.

\begin{figure}
\includegraphics[width=8.8cm]{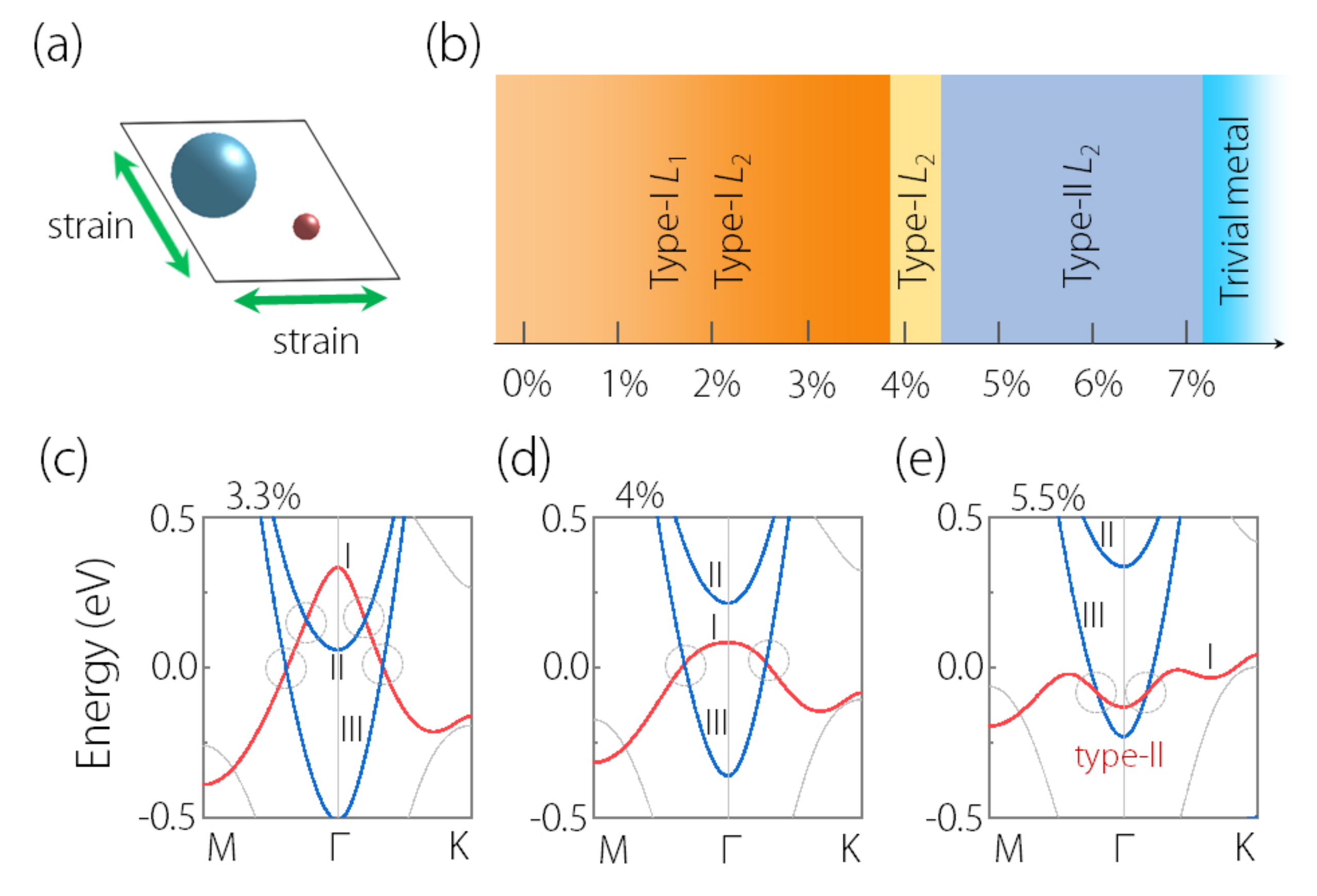}
\caption{(a) Schematic figure of K$_2$N monolayer under biaxial in-plane strain. (b) Nodal line phase diagram of K$_2$N monolayer under biaxial in-plane strain. Electronic band structures of K$_2$N along the $M$-$\Gamma$-$K$ path under (c) 3.3$\%$, (d) 4.0$\%$, and (e) 5.5$\%$ tensile strains.
\label{fig6}}
\end{figure}

Here we show the case of applying biaxial in-plane strains on K$_2$N monolayer, as shown in Figs.~\ref{fig6}(a). It is worth noticing that, all the crystal symmetries including $M$$_z$ will preserve under biaxial in-plane strains. As shown in Figs.~\ref{fig6}(b), a rich phase diagram can be realized in K$_2$N monolayer by such strain engineering. Under small tensile strains ($< 3.8$$\%$), we find both nodal lines do not change much. Fig.~\ref{fig6}(c) shows the band structure under a 3.3$\%$ tensile strain. We find the crossings between band I and band II (band III) retain, which produce two type-I nodal lines (type-I $L$$_1$ and type-I $L$$_2$). Under tensile strains in the range of 3.8$\%$-4.3$\%$, we find $L$$_1$ will be annihilated but $L$$_2$ retains. Figs.~\ref{fig6}(d) shows the band structure under a 4.0$\%$ tensile strain. We find band I is separated from band II, while the crossing between band I and band III is retained. As the result, K$_2$N monolayer only shows a type-I nodal line ($L$$_2$) in this case. Very interestingly, we find nodal line $L$$_2$ can further transformed into a type-II nodal line under much larger tensile strains (4.3$\%$-7.1$\%$). The corresponding band structure is shown in Figs.~\ref{fig6}(e). When apply tensile strains larger than 7.1$\%$, the crossings between band I and band III are also annihilated. As the results, K$_2$N monolayer becomes a trivial FM metal [see Figs.~\ref{fig6}(b)]. These results indicate K$_2$N monolayer can serve as an excellent platform to investigate novel nodal line transformations in time-reversal symmetry breaking 2-D system.

\section{Discussion and conclusion}
Before closing, we have several remarks. First, the most important finding here is that, we propose the first example of 2-D nodal line semimetal in a \emph{$d^0$} ferromagnet and the nodal lines are robust against SOC. To our best knowledge, all the FM 2-D nodal lines proposed so far are identified in \emph{d/f}-type ferromagnets. Comparing with \emph{d/f}-type counterpart, \emph{$d^0$} ferromagnets are promising to show higher carrier mobility and longer spin coherence length because of the much weaker localization of $p$ electrons and the smaller SOC strength~\cite{47,48,49}. This can bring significant advantages for high-speed and long-distance transports. Moreover, 2-D nodal line materials robust against SOC are also very rare, which is only proposed half metal MnN monolayer previously~\cite{40}. In K$_2$N monolayer, we find the nodal lines are strictly protected by symmetry and are also robust against SOC.

Second, K$_2$N monolayer can realize a room temperature ferromagnetism with the Curie temperature as high as 942 K. In Figs.~\ref{fig2}(d), we compare the Curie temperature between K$_2$N monolayer and other known FM 2-D nodal line materials. We can clearly observe that, the Curie temperatures for the examples proposed previously are almost lower than the room temperature, which greatly hinders their future applications in spintronic devices. However, K$_2$N monolayer has a significantly higher Curie temperature than them.

Finally, the nodal lines in K$_2$N monolayer also show several additional features: (i) the system preserves the inversion symmetry but lacks the time-reversal symmetry, thus the nodal lines are time-reversal breaking Weyl lines; (ii) the linear dispersion range for both nodal lines are very large ($> 1$ eV); (iii) one of the nodal lines can transform from type-I to type-II nodal line under specific lattice strains [see Figs.~\ref{fig6}(b) and (e)], which makes topological properties of K$_2$N monolayer even meaningful; (iv) the nodal line band structures are fairly clean without coexisting of other extraneous bands, which greatly favors their detections in experiments.

In conclusion, we have demonstrated K$_2$N monolayer as the first 2-D Weyl nodal line semimetal in \emph{$d^0$} ferromagnets. We show that K$_2$N monolayer is both thermally and dynamically stable. The material is a \emph{$d^0$} ferromagnet with the magnetism arises from the partially filled $p$ orbitals. The material has a Curie temperature of 942 K, which is among the highest in known 2-D FM materials. In the low energy band structures, the material shows two concentric Weyl nodal lines in the 2-D Brillouin zone. Both nodal lines are under the protection of mirror symmetry and can retain under SOC. We have built an effective model, which can well characterize the emergence of nodal lines in the material. In addition, we find the material exhibits rich nodal line transformation under biaxial in-plane strains, where a single type-I or type-II nodal line can be realized. Our work provides a candidate material to investigate topological nodal line states in 2-D \emph{$d^0$} ferromagnet and has potential applications in spintronic devices.

\begin{acknowledgments}
This work is supported by National Natural Science Foundation of China (Grants No. 11904074), Nature Science Foundation of Hebei Province (No. E2019202222 and E2019202107). One of the authors (X.M. Zhang) acknowledges the financial support from Young Elite Scientists Sponsorship Program by Tianjin.
\end{acknowledgments}

\end{document}